\documentclass[twocolumn, a4paper]{revtex4}
\usepackage{graphicx}
\usepackage{dcolumn}
\usepackage{amsmath}
\begin{document}
\hsize\textwidth\columnwidth\hsize\csname@twocolumnfalse\endcsname

\title{Direct measurement of the maximum tunnel rate in a radio frequency single electron transistor operated as a microwave mixer}
\author{D. J. Reilly\footnote{Electronic mail:djr@phys.unsw.edu.au} and T. M. Buehler}
\affiliation{Centre for Quantum Computer Technology, School of Physics, University of New South Wales,
Sydney 2052, Australia}

\begin{abstract}
By operating the radio frequency single electron transistor (rf-SET) as a mixer we present  
measurements in which the $RC$ roll-off of the tunnel junctions is observed at high frequencies. Our technique makes use of the non-linear rf-SET transconductance to mix 
high frequency gate signals and produce {\it difference-frequency} components that fall within the bandwidth of the rf-SET. At gate frequencies $>$15GHz the induced charge 
on the rf-SET island is altered on time-scales faster than the inverse tunnel rate, preventing mixer operation. We suggest the 
possibility of utilizing this technique to sense high frequency signals beyond the usual rf-SET bandwidth.
\end{abstract}

\maketitle

% Intro

The tunneling of single electrons in metallic nano-scale junctions is a complex process involving the undressing and dressing of 
quasi-particles as they tunnel through an insulating barrier \cite{natobook}. In recent times the specific details of this process have received considerable attention in 
relation to junctions in the Coulomb blockade (CB) regime, with a charging energy greater than the energy of thermal fluctuations ($E_C > kT$) and a tunneling resistance 
exceeding the resistance quantum ($R_T > R_K = h/e^2 \sim 26 $k$\Omega$) \cite{Averinlikharevbook}. Such junctions are
characterized by three time-scales \cite{natobook}. The shortest of these is the tunneling
time which can be related to the time that the electron spends under the barrier \cite{Hauge_Stovneng_RMP_89}. For typical nano-scale junctions in the CB regime this time 
is exceedingly short ($<10^{-15}$s and assumed to be zero in the orthodox theory \cite{Averinorthodox}). In the intermediate 
regime, virtual tunneling processes are subject to the uncertainty principle which defines a time-scale that is connected with the 
Coulomb charging energy, $\tau = \hbar/2 \pi E_C$. The longest characteristic time-scale is set by the tunnel resistance $R_T$ and junction capacitance $C_J$. This 
time-scale is the reciprocal of the tunnel rate for a junction biased at the threshold voltage for CB ($V_{sd}=e/C_J$). Here we present a technique that 
directly measures this maximum tunnel rate, ($f_{max}=I/e\sim1/8 R_T C_J$) in a Al/AlO$_x$ radio frequency  
single electron transistor (rf-SET) \cite{Schoelkopf_science} by operating the device as a heterodyne-mixer of microwave charge signals. 

Previously, Knobel {\it et al.,} \cite{Knobel_APL} demonstrated the use of the single electron transistor (two tunnel junctions in series) as a {\it radio-frequency} mixer 
by making use of the non-linear dependence of source-drain current on gate charge. Here we extend this concept to microwave frequencies and access the regime beyond the intrinsic tunnel rate of the SET 
tunnel junctions. Contrasting Knobel {\it et al.,} our experiment makes use of the rf-SET which consists of a SET embedded in a $LCR$ impedance 
matching `tank' circuit. At the 
resonance frequency ($f_{tank}$) the tank circuit 
transforms the high impedance of the SET (typically 50k$\Omega$) towards 50$\Omega$. With the SET and tank circuit terminating a 50$\Omega$ transmission line, a rf carrier 
signal is introduced so that the amount of 
reflected power depends on the resistance of the SET which in turn is a function of the induced charge on the SET `island'. In comparison to mixing with conventional dc 
SETs, the rf-SET is advantageous since the high bandwidth (typically $>$10MHz) 
translates to a large intermediate frequency (IF) bandwidth allowing operation well above 
$1/f$ noise where rf-SETs have demonstrated near quantum-limited charge sensitivities \cite{Aassime_APL, Buehler_JAP04}. Similar to rf-SET based cryogenic amplifiers, 
rf-SET mixers also exhibit low power dissipation, low capacitance and an extremely large input impedance \cite{Segall_APL}. 

In the experiments presented here we focus on Al/AlO$_x$ SETs formed by a standard shadow mask evaporation process \cite{dolan}. Our devices typically have 
charge sensitivities of better than $\delta q=10^{-5}e/\sqrt{Hz}$ and a maximum 
tunnel rate close to 16GHz $\sim 1/2 \pi R_{\Sigma} C_{\Sigma}$, for a total resistance $R_{\Sigma}$=50k$\Omega$ and total capacitance $C_{\Sigma}$=0.2fF. This $RC$ time 
constant sets a limit on the rate at 
which the SET 
island 
can be charged or discharged by a single electron. This {\it intrinsic} bandwidth however, is significantly larger than the bandwidth of the measurement 
electronics (kHz for conventional SETs or up to 100MHz for the rf-SET) so that such 
effects are usually not observed in ordinary transport experiments. 

In order to circumvent this problem of limited bandwidth and get to the intrinsic limit set by the 
tunnel barriers we operate the rf-SET as a mixer, utilizing the inherent non-linear dependence of the  source - drain current $I_{sd}$ on gate voltage $V_G$. In this 
regime, 
the SET will {\it multiply} gate signals of different frequencies ($f_{sig1}$ and $f_{sig2}$) and the current $I_{sd}$, will contain mixing 
terms at the sum and difference of these frequencies, $f_{dif}=f_{sig1}-f_{sig2}$. Although the frequencies of the gate signals may exceed the bandwidth of the rf-SET 
(10MHz in our case), the rf-SET will continue 
to operate as a mixer so that the {\it difference} frequency can be chosen to fall within the rf-SET bandwidth: $f_{dif}<$10MHz. The detection of this difference term 
is possible until the gate signals induce charges on the island at a rate faster than the maximum tunnel rate of the tunnel junctions. At gate frequencies 
beyond the maximum tunnel rate, the amplitude of the difference component will approach zero since electrons are unable to charge (or discharge) the island and 
current flow from source to drain is suppressed. We note that the rf-SET can also exploit the non-linearity between $I_{sd}$ and source-drain voltage $V_{sd}$ to mix 
signals in a 
similar way to superconducting mixers \cite{Tucker_RMP}. 
%Figure 1
\begin{figure}[t!]
\begin{center}
\vspace{0.5cm}
\includegraphics[width=7.0cm]{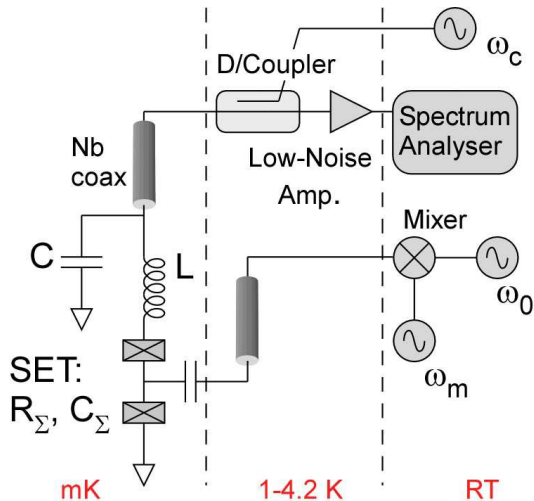}
\caption{Experimental configuration in which the rf-SET is operated as a mixer. The room temperature (unbalanced) mixer produces the three frequencies that are fed to a 
single gate. The amplitude of the {\it beating} frequency is measured with a spectrum analyzer after amplification by a low-noise cryogenic 
amplifier.}   
\vspace{-0.5cm}
\end{center}
\end{figure}

Turning to the details of our measurement technique (see Figure 1), we bias the SET to a non-linear region of the transconductance and apply an amplitude modulated (AM) 
waveform to the gate electrode by mixing the output of a microwave generator ($\omega_0$) with a 200kHz ($\omega_m$) local oscillator. The (unbalanced) mixer produces 
three frequencies at $\omega_0,(\omega_0+\omega_m)$ and $(\omega_0-\omega_m)$ which are all fed to a single gate electrode. While fixing the modulation frequency 
$\omega_m$=200kHz, we sweep the frequency of the microwave signal $\omega_0$ from 1GHz to 27GHz. The SET non-linear transconductance will now mix these three 
frequencies with each other to produce a 
current $I_{sd}$ containing sum and difference components at  $\omega_m$, $-\omega_m$, $(2\omega_0-\omega_m)$, $(2\omega_0+\omega_m)$ and $2\omega_m$, as well as higher 
order terms. We focus on the difference component at $\omega_m$, which corresponds to the fixed microwave modulation 
frequency  (200kHz) and  falls within the bandwidth of the rf-SET and above the corner frequency for $1/f$-type charge noise \cite{Schoelkopf_science}. To first order, the 
amplitude of $\omega_m$ depends linearly on the amplitude of the high frequency (microwave) currents induced by the gate signals.
Finally for the case of the rf-SET, where the gate signals are mapped to changes in reflected power, we include the effect of the rf carrier signal $\omega_c$ rather 
than current $I_{sd}$. In this case the gate signals will modulate $\omega_c$ \cite{Aassime_APL,Buehler_JAP04} and produce side-bands 
at frequencies now shifted upwards by $\omega_c$, so that $\omega_m$ (similar for all other components) is now shifted to $\omega_c+\omega_m$. In the discussion that 
follows we refer to the component at $\omega_c+\omega_m$ as the {\it difference} (or beating) frequency.

Figure 2 shows the reflected 
rf power signal from the rf-SET as a function of frequency. For the purpose of illustrating the technique we set $\omega_m$=10kHz, and $\omega_0$=100kHz so that {\it all} 
components fall within the 10MHz bandwidth of the rf-SET and can be displayed with a spectrum analyzer. Note the difference component $\omega_m$ symmetric about the rf 
carrier $\omega_c$ at  
$\omega_c\pm\omega_m$=328.21MHz and 328.19MHz.  Increasing the frequency of the microwave gate signal $\omega_0$ beyond the 10MHz bandwidth of the rf-SET (where it 
cannot be displayed) does not alter the frequency of the difference component at $\omega_c+\omega_m$, although its amplitude is 
directly related to the amount of current that flows at the modulated frequency $\omega_0$.    

%Figure 2
\begin{figure}[t!]
\begin{center}
\vspace{0.5cm}
\includegraphics[width=7.5cm]{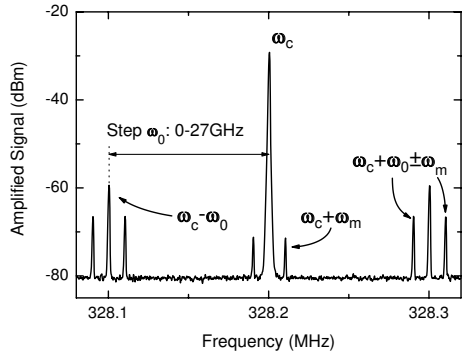}
\caption{Frequency spectrum obtained with a spectrum analyzer for the case where $\omega_0$=100kHz, $\omega_m$=10kHz and $\omega_c$=328.2MHz. For illustrative purposes 
these frequencies have been chosen to fall within the bandwidth of the rf-SET tank circuit. Note that increasing the frequency $\omega_0$ does not affect the frequency of
$\omega_c+\omega_m$.} 
\vspace{-0.5cm}   
\end{center}  
\end{figure}  

To verify our ability to detect the beating frequency of microwave gate signals, we now set $\omega_0$ to $\sim$8GHz and modulate this signal at $\omega_m$=200kHz. Figure 
3a) shows the rf-SET response (different device to Fig.2) where $\omega_c$=278.25MHz and the difference frequency $\omega_c+\omega_m$ is at 278.45MHz. As an indication 
that this component ($\omega_c+\omega_m$) is due to the SET acting as a mixer we operate the SET in the superconducting state ($B$=0) and change 
$V_{sd}$ to vary the non-linear transconductance. At $V_{sd}$=0, the SET is biased to the middle of the superconducting gap and the mixing response drops to zero. At 
$V_{sd}$=0.75mV the SET is biased to the threshold for quasi-particle transport and the amplitude of the component at $\omega_c+\omega_m$ reaches a 
maximum in 
connection with the non-linear transconductance. Although we focus on rf-SETs in the superconducting state where the charge sensitivity reaches a maximum 
\cite{Schoelkopf_science}, similar behavior was also observed for normal state ($B$=1T) devices. Figure 3b) shows the gain dependence of $\omega_c+\omega_m$ as a function 
of dc-offset charge applied to the gate. In the upper panel we show CB oscillations and indicate the regions of high non-linear transconductance that correspond to the 
peaks in amplitude of $\omega_c+\omega_m$ as indicated in the lower panel. 

%Figure 3
\begin{figure}[t!]
\begin{center}
\vspace{0.5cm}
\includegraphics[width=7.5cm]{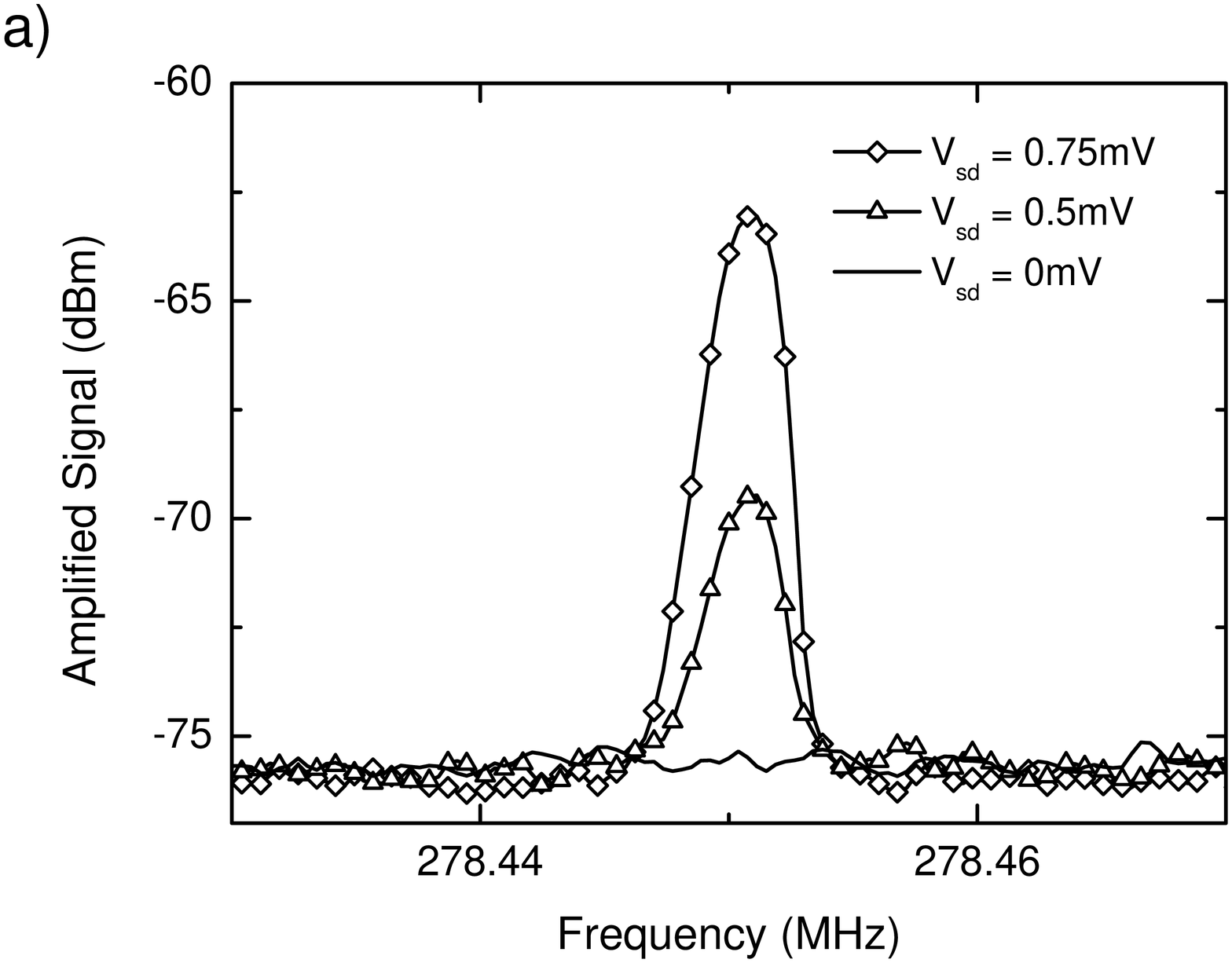}
\includegraphics[width=7.5cm]{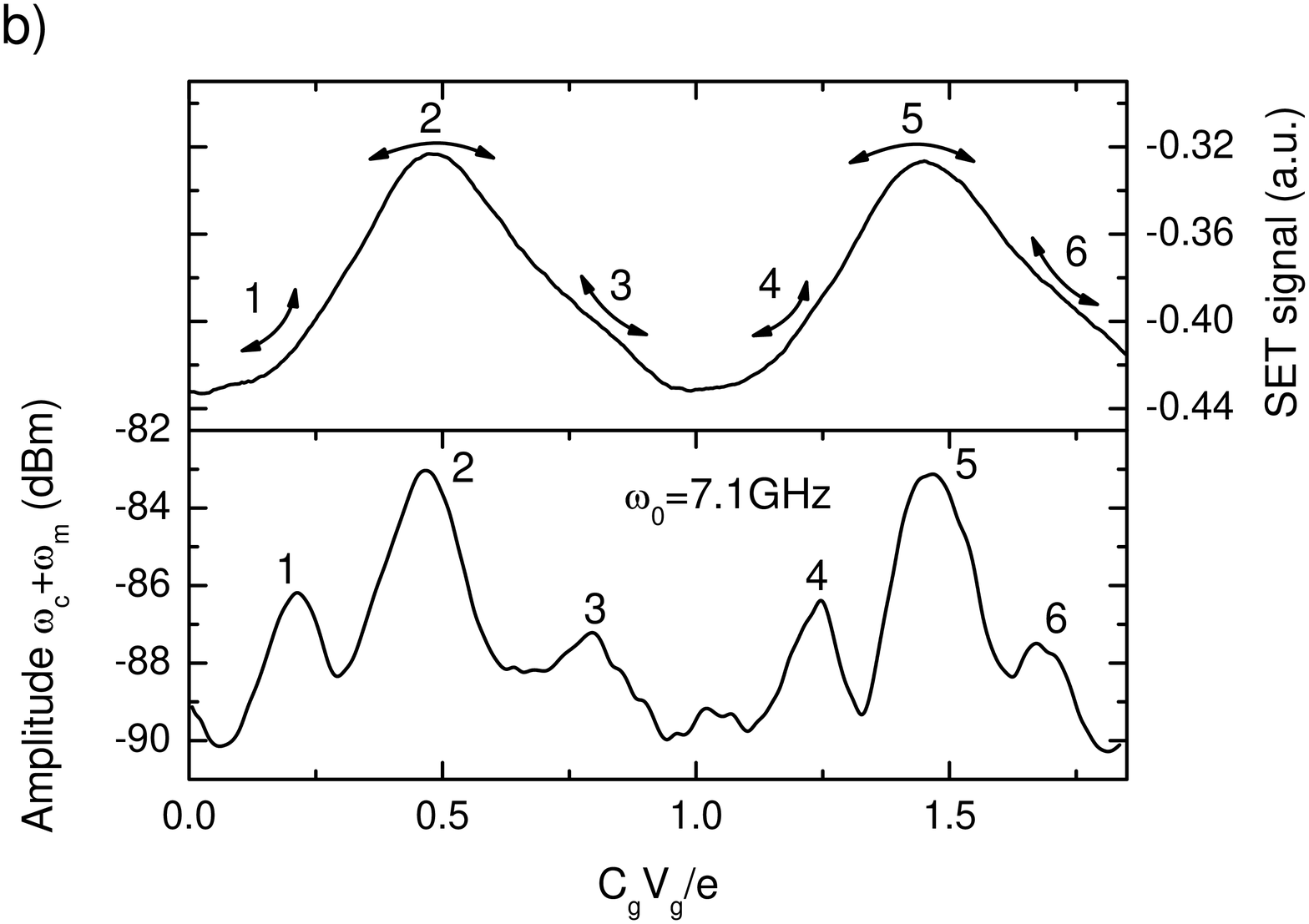}
\caption{{\bf a)} Amplitude dependence of $\omega_c+\omega_m$ for a rf-SET in the superconducting state. At $V_{sd}$=0mV, the SET is biased to the superconducting gap so 
that there is no non-linearity available for mixing. The beating signal at $\omega_c+\omega_m$ reaches a maximum amplitude with the SET biased to
$V_{sd}$=0.75mV, the onset of quasi-particle tunneling. {\bf b)} Gain dependence of $\omega_c+\omega_m$ as a function of dc-offset charge $C_gV_g/e$. Upper panel shows CB 
oscillations where the numbered arrows indicate regions of non-linear transconductance. Lower panel shows corresponding amplitude dependence of $\omega_c+\omega_m$.}
\vspace{-0.5cm}
\end{center}
\end{figure}

%Figure 4
\begin{figure}[t!]
\begin{center}
\vspace{0.5cm}
\includegraphics[width=7.5cm]{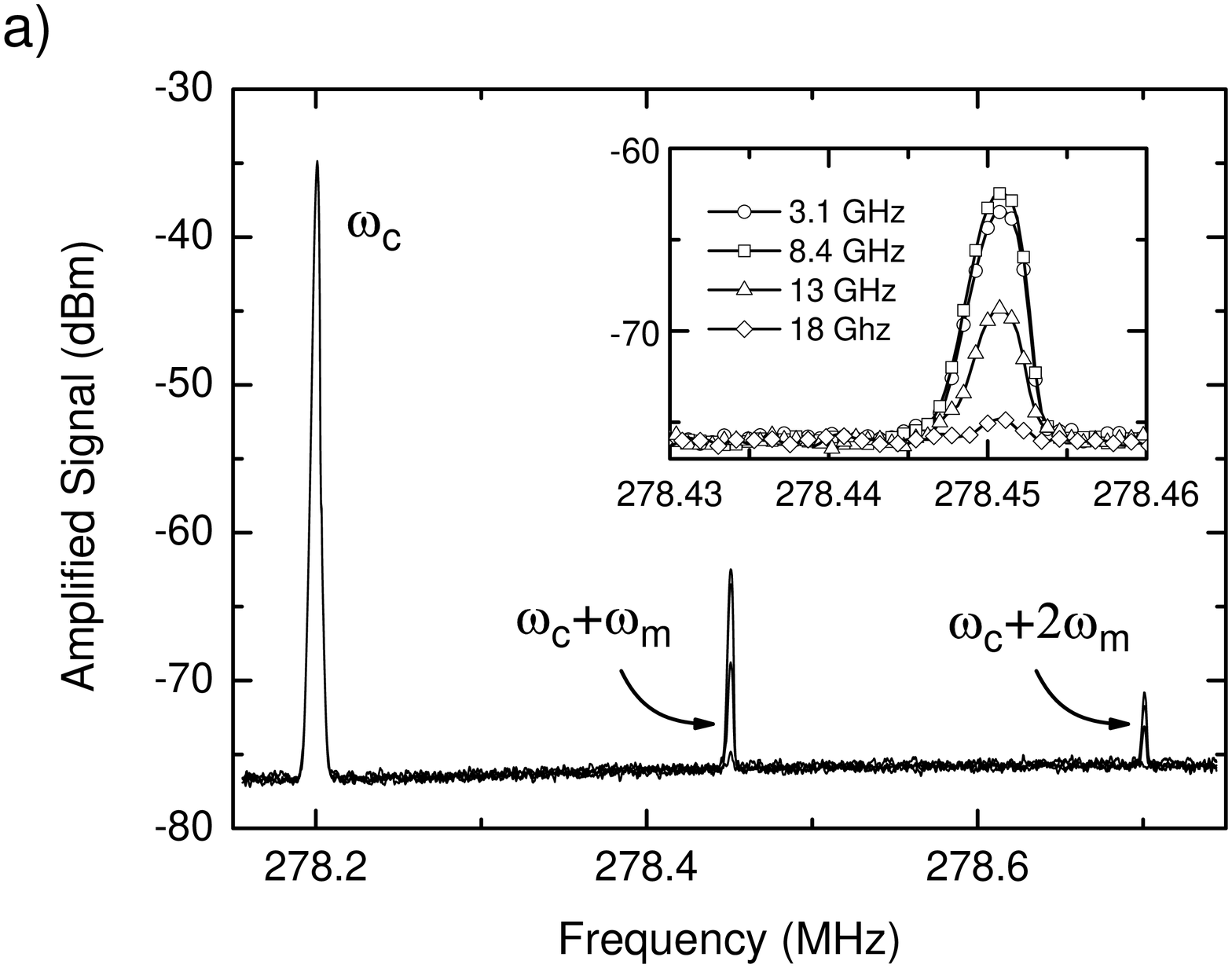}
\includegraphics[width=7.5cm]{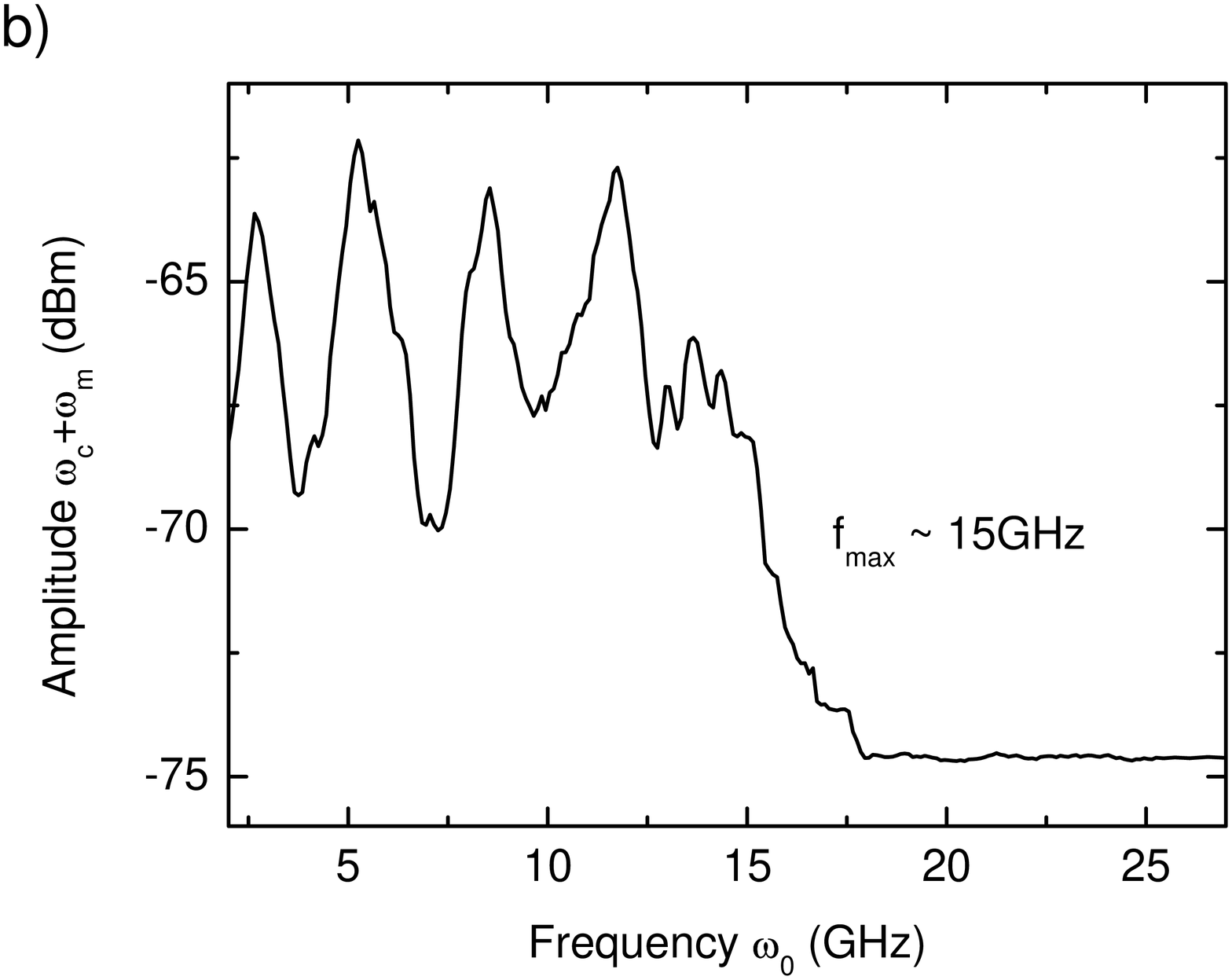}
\caption{{\bf a)} Spectrum showing the carrier peak $\omega_c$ and the amplitude dependence of $\omega_c+\omega_m$ and $\omega_c+2\omega_m$ as a function of the frequency
of $\omega_0$, $B$=0, $V_{sd}$=0.75mV. Inset: is a zoom of the $\omega_c+\omega_m$ peak. {\bf b)} shows the amplitude of the side-band at
$\omega_c+\omega_m$ as a function of the frequency of $\omega_0$ (5pt average). The reduction in amplitude of the beating frequency is associated
with a decrease in the high frequency current as the island potential is modulated at a rate exceeding the maximum tunnel rate.}
\vspace{-0.5cm}
\end{center}
\end{figure}

Turning now to our key result, Figure 4 shows the response spectra obtained for a superconducting rf-SET ($V_{sd}$=0.75mV) as $\omega_0$ is increased to 
27GHz. 
In addition to the difference component at $\omega_c+\omega_m$=278.45MHz (Figure 4a) note the presence of the component at $\omega_c+2\omega_m$=278.70MHz, which is 
associated with the difference frequency between 
the gate signals, $(\omega_0-\omega_m) - (\omega_0+\omega_m)$ = $2\omega_m$. The amplitude of these mixing 
components remains finite (with some degree of oscillation) as a function of $\omega_0$ until $\sim$14GHz, after which both $\omega_c+\omega_m$ and 
$\omega_c+2\omega_m$ are steadily reduced. 
Figure 4b shows the amplitude of $\omega_c+\omega_m$ as a function of $\omega_0$, where the data has been smoothed using a 5 point average. We first focus on the 
characteristic roll-off in signal as $\omega_0$ is increased above $\sim$15GHz. We attribute this steady decrease in amplitude to a reduction in current 
at frequencies $\omega_0$ and $\omega_0+\omega_m$ in connection with changes in the SET island potential at a rate beyond the maximum tunnel rate of the junctions.  
Independent measurements of $C_\Sigma$ and $R_\Sigma$ are consistent with a maximum frequency near 15GHz ($1/2 \pi R_{\Sigma} C_{\Sigma}\sim$15GHz, for a differential 
resistance $R_{\Sigma}=25$k$\Omega$ at $V_{sd}$=0.75mV and $C_{\Sigma}=0.46$fF). Additional measurements on other SETs confirm this behavior, with the roll-off frequency 
controlled by  the differential resistance, $R_{\Sigma}$ and $C_{\Sigma}$. At gate frequencies $\hbar\omega_0>>kT$, we also observe the signatures of photon 
assisted tunneling \cite{Tien_Gordon, Fitzgerald}. These effects are only evident when the rf-SET is not operating as a mixer and do not seem to alter  
our measurement of the maximum tunnel rate. 

The oscillations in amplitude of  $\omega_c+\omega_m$ at $\omega_0<$ 15GHz are particularly interesting. At present the exact mechanism responsible for 
this dependence is unclear, although we note that oscillations are always present, (but qualitatively different for different SETs) and sensitive to the SET bias, 
$V_{sd}$. The most plausible explanation is that the non-monotonic dependence is due to the microwave gate signals coupling to the source-drain leads of the SET 
and higher-order resonances in the tank circuit. Parasitic capacitances and inductances associated with bond-wires and 
lithographically defined connector pads are likely to produce additional resonances at frequencies beyond the 
conventional first-order tank resonance at $\omega_c$=278.2MHz. Gate signals coupling to these resonances could disturb the rf-response of the carrier power and depend on 
the SET resistance controlled by $V_{sd}$. 

Using a similar method to the measurement technique presented here, we suggest that the rf-SET-mixer may also be useful in detecting charge noise at frequencies beyond 
the tank circuit bandwidth. In such a measurement a microwave local-oscillator (LO) gate signal is applied to 
`down-convert' high frequency charge noise to the tank circuit `base-band' for integration over the 10MHz bandwidth. Increasing the LO frequency in 10MHz steps would 
enable the noise power to be measured up to the cut-off frequency defined by the maximum tunnel rate. Such an experiment would be advantageous in studying 
frequency-dependent shot noise 
\cite{shoelkopfshotnoise,engelloss} and the magnitude of noise associated with rapidly switching electron traps \cite{Kautz_PRB_00,Buehler_RTSJAP}. 
Finally we draw attention to the possibility of using the rf-SET-mixer to detect and readout an oscillating charge dipole. Although significant further work is 
needed to explore this possibility, such a scenario is of considerable interest for performing quantum non-demolition measurements \cite{grangier,wallraff} of 
(weakly coupled) charge qubits. 

In conclusion, we have presented a technique in which the rf-SET is operated as a mixer to directly measure the maximum tunnel rate set by the resistance and capacitance 
of the SET tunnel junctions. Our results are consistent with the rate expected from an indirect measure of $C_{\Sigma}$ and $R_{\Sigma}$ and suggest the possibility 
of using this technique to detect high frequency charge signals beyond the usual bandwidth set by the rf-SET tank circuit. 
  
The authors would like to thank A. J. Ferguson, R. J. Schoelkopf, K. Schwab, A. Korotkov and R. Brenner for many useful and insightful conversations.
We also thank R. G. Clark, A. R. Hamilton, A. S. Dzurak, R. P. Starrett and D. Barber for helping to make these experiments possible. 
This work was supported by the  Australian Research Council, the Australian Government and
by the US National Security Agency (NSA), Advanced Research and
Development Activity (ARDA) and the Army Research Office (ARO)
under contract number DAAD19-01-1-0653. DJR acknowledges a
Hewlett-Packard Fellowship.
\vspace{-0.5cm}
\small{

}
\end{document}